# Modeling of Liquid Water on CM Meteorite Parent Bodies and Implications for Amino Acid Racemization

**Barbara A. Cohen**
Department of Planetary Sciences, The University of Arizona, Tucson AZ 85721
E-mail: bcohen@lpl.arizona.edu

and

**Robert F. Coker**
Department of Physics, The University of Arizona, Tucson AZ 85721
(now at the Department of Physics and Astronomy, University of Leeds, Leeds LS2 9JT, UK)

## Abstract

We have constructed an asteroid model with the intent of tracking the radial and temporal dependence of temperature and composition throughout a 100-km diameter CM-type parent body, with emphasis on constraining the temperature and duration of a liquid water phase. We produce a non-uniform distribution where liquid water persists longest and is hottest in the deepest zones and the regolith never sees conditions appropriate to aqueous alteration. We apply the model predictions of the liquid water characteristics to the evolution of amino acids. In some regions of the parent body, very little change occurs in the amino acids, but for the majority of the asteroid, complete racemization or even destruction occurs. We attempt to match our thermal model results with CM meteorite observations, but thus far, our model does not produce scenarios that are fully consistent with these observations.

## Introduction

Carbonaceous chondrites, especially CM meteorites, are known to contain extraterrestrial organic material in an aqueously altered matrix (Cronin *et al*. 1988). Included in meteoritic organic material are over 80 different amino acids, which are protein building blocks. The majority of amino acids used by Earth organisms are left-handed (L-enantiomers) rather than right-handed (D-enantiomers). Recently, Cronin and Pizzarello (1997) showed that some meteoritic amino acids may have an excess of L-enantiomers (7-9%). Organic material delivered to Earth through meteorite bombardment may have contributed to life's preference for left-handed amino acids.

Cohen and Chyba (1999) explored the conditions under which some easily-racemized meteoritic amino acids could retain an enantiomeric excess throughout their lifetime. These types of calculations depend strongly on the characteristics of the aqueous phase on the asteroid parent body, particularly its duration and temperature. Previous thermal modeling of carbonaceous chondrite parent bodies (DuFresne and Anders 1962, Grimm and McSween 1989) has characterized the conditions that produce a liquid water phase on a parent asteroid, but has been less useful at being able to describe this phase in detail.

This work focuses on specifying the temperature and duration of the liquid water phase on asteroid parent bodies. The CM meteorite characteristics have been chosen in order to use this work to model organic evolution and survival in a parent body such as that which produced Murchison.

## CM Meteorite Characteristics

CM chondrites are carbon-rich, aqueously altered, stony meteorites. CM meteorites are interesting because up to 30% of their carbon content exists as complex organic material (Rubin 1998). A number of other carbonaceous meteorites have also proved to contain organic material of extraterrestrial origin, but the Murchison-type CM meteorites contain the most (Cronin *et al*. 1988). The conditions on the CM parent body can be constrained both by the presence of organic material and by the unique CM mineralogy.

The CM meteorites are extensively altered, but some still contain minichondrules and some relic olivine and pyroxene crystals (McSween 1979). The primary minerals are serpentine-group minerals, comprising 55-85% of the meteorite by volume (Scott *et al*. 1988). Other important minerals are the layered hydroxysulfide tochilinite, magnetite and pyrrhotite, carbonates, and soluble Mg salts (Zolensky and McSween 1988). This suite of minerals is indicative of aqueous alteration, at least some of which occurred on the asteroid parent body as opposed to in the solar nebula itself (Bunch and Chang 1978, 1980). The conditions of aqueous alteration have been constrained by a number of different methods, including mineral stability relationships (Zolensky 1984) and isotope equilibration arguments (Rowe *et al*. 1994, Clayton and Mayeda 1984, Lerner 1995). It is generally thought that the aqueous alteration occurred over short ($10^3$-$10^4$ yr) timescales at cool (273-300K) temperatures. Previous modeling (DuFresne and Anders 1962, Grimm and McSween 1989, Zolensky *et al*. 1989) has produced scenarios that are consistent with these estimates.

## The Model

The FORTRAN code written for this purpose solves the one-dimensional radially symmetric heat conduction equation for an asteroid 100 km in diameter.

The model asteroid has seven components: $H_2O$ ice, $H_2O$ liquid, void space filled with $H_2O$ vapor, Mg-olivine (forsterite), Mg-pyroxene (enstatite), hydrated rock (antigorite serpentine), and non-reactive rock (representing the ~40% of CM meteorites that is not serpentinized). We realize that this is an extreme simplification of the mineralogy and that neglected phases such as carbonates, sulfides, iron metal, and iron oxides could have important effects on the nature of the fluid-rock interactions. However, the lack of data on the kinetics of basic serpentinization (discussed below) makes further refinement of the chemical reactions impossible. We do include salts to the extent that the freezing point of liquid water in the model is set to saturated $MgSO_4$ solution properties.

The starting proportions of each component are varied to force the final composition to match CM meteorite observations. All components are assumed to have particle sizes on the order of tens of micrometers, the particle size of the lunar regolith (Heiken *et al*. 1991) and are uniformly distributed throughout each zone and throughout the entire asteroid. The state of $H_2O$ as it is accreted is determined by the initial material temperature $T_i$ (see Eq. A3). Thus far, we have only considered scenarios where ice accretes.

We assume that the timescale for accretion is rapid, but solar nebular models (Cassen 1994, Wood and Morfill 1988) indicate that the nebula was too turbulent for substantial accretion until about $10^6$ years after the molecular cloud collapse formed a disk around the sun. We are using the collapse time here to also indicate the formation of radionuclides; therefore, an "accretion time" of 1 Ma indicates that radioactive material is allowed to decay from its starting abundance (as detailed in the Appendix) for 1 Ma before being incorporated into the asteroid.

Our model incorporates the solar nebular temperature variability predicted by Cassen (1994) to calculate the temperature of material being accreted and the surface equilibrium temperature of the body throughout the lifetime of the nebula. We recognize that this is not the only solar nebula model in existence, but it is the only one with a quantitative temperature-time-distance relationship. We discuss the effects of higher-temperature nebula models below. After $10^7$ years, the nebula would dissipate and the asteroid surface would be in solar equilibrium with an evolving main-sequence sun (Endal and Sofia 1981), but our calculations do not go that far in time.

Heat is contributed by accretion and decay of both short- and long-lived radionuclides, dominated by $^{26}$Al. We vary the time between nebular collapse and the accretion of the asteroid, which dictates the amount of live radionuclides. Because of the rapid decay of these nuclides, the aqueous phase must begin early on the parent body, and last a relatively short time. Solar inductive heating has been proposed as an alternative heat source in asteroids (Sonnett *et al.* 1970, Herbert 1989). However, due to the largely unconstrained nature of the early solar magnetic field and CM meteorite electrical conductivity values, we do not consider inductive heating in our model at this time.

Thermal properties of the materials are composition- and temperature-dependent where appropriate. We include provisions for convection, hydraulic fracturing, and vapor diffusion. The aqueous alteration found in carbonaceous chondrites is represented by the formation of antigorite (Mg-serpentine) via the reaction:

$$\mathrm{Mg_2SiO_4\ (forsterite) + MgSiO_3\ (enstatite) + 2\ H_2O \rightleftarrows Mg_3Si_2O_5(OH)_4\ (antigorite)}$$

proceeding only in the presence of liquid H$_2$O with a pressure- and temperature-dependent rate of reaction determined for a similar reaction by Wegner and Ernst (1983).

**TABLE I**

**Summary of parameters varied in model runs**

| Run number | Accretion time (Ma after nebula collapse) | Heliocentric distance (AU) | Permeability | Volume fraction of ice accreted |
|---|---|---|---|---|
| 1 | 3 | 3 | medium | 30 |
| 2 | 3 | 3 | medium | 20 |
| 3 | 3 | 3 | medium | 40 |
| 4 | 3 | 3 | low | 30 |
| 5[1] | 3 | 3 | high | 30 |
| 6[2] | 3 | 3.5 | medium | 30 |
| 7[3] | 2 | 3 | medium | 30 |
| 8[2] | 2 | 3 | medium | 40 |
| 9[3] | 2 | 3.5 | medium | 30 |
| 10 | 3.5 | 3 | medium | 30 |

1. Terminated at ~6 Ma (due to computational limits) after most of the water has frozen out or reacted.
2. Terminated at less than 10 Ma but after water at all radii has frozen out or reacted.
3. Terminated at ~3.5 Ma when central temperature exceeded 600 K.

# Modeling Results

Table I is a summary of the parameter space we explore in 10 runs. For all runs, we use an initial volumetric composition of 22% forsterite, 17% enstatite, 16% void, and the tabulated percentage of ice. The remaining volume is non-reactive rock. This choice of mineralogy forces the final assemblage (if fully hydrated) to become 60% serpentine, have a small amount of olivine and pyroxene left over, and have a bulk porosity of 20%. The bulk porosity then matches measured bulk porosities of Murchison (Britt and Consolmagno 1997, Corrigan *et al.* 1997) and the ~5% porosity created from the serpentinization also matches well with measured CM matrix porosities (Corrigan *et al.* 1997).

Run 1 uses our "canonical" parameters: accretion 3 Ma after nebula collapse, formation 3 AU from the sun, permeability of $10^{-12}\,m^2$, and an initial ice fraction of 30% by volume. Fig. 1 shows that, for Run 1, the peak temperature at the center of the asteroid is 380K, and the temperature structure of the asteroid 1 Ma after it accretes (4 Ma after nucleosynthesis) is shown in Fig. 2. Throughout the asteroid, liquid water can persist for over $10^6$ years (Fig. 3). Fig. 4 shows the temperature and duration of liquid water near the center of the asteroid. In Run 1, liquid water exists for about 2.3 million years, peaking at 375K. The remaining runs are compared to this canonical run.

*Ice Fraction:* The initial amount of ice (if it is not the limiting reagent) has the simplest effect on the temperature and duration of the liquid phase. The higher ice fraction (Run 3) results in a slightly lower maximum central temperature (360K) (Fig. 1). The surplus of water does not have enough time to diffuse away from the center and freezes in the central zones after existing for 3.5 Ma (Fig. 4). The lower ice fraction (Run 2) means that water is nearly the limiting reagent, so most of it reacts with the mineral fraction and the small remainder diffuses out in another $5 \times 10^3$ years (Fig. 4), with a higher maximum temperature (390K) (Fig. 1). Because diffusion moves water away from the reacting region, a smaller initial volume of water means a potentially shorter duration of liquid water in any particular zone. For example, in Run 3, the water freezes before all of it has had a chance to diffuse, and so ice will remain in the central regions.

*Permeability:* Water vapor carries heat from the interior of the asteroid by diffusing through the rock. A low permeability of $10^{-13}$ m$^2$, which is appropriate for a fully lithified rock such as sandstone (Turcotte and Schubert 1982), prevents heat loss by vapor transport. In this case (Run 4), liquid water has a peak temperature of 390K and persists for ~4.2 Ma before diffusing outward (Fig. 4). A high permeability of $10^{-11}$ m$^2$, which is appropriate for sandy soils like the lunar regolith (Heiken *et al.* 1991), allows the center to cool though vapor transport. In Run 5, the peak temperature is lower, about 370K, and water diffuses out of the central region after 1.5 Ma (Fig. 4). Clearly, a higher permeability is conducive to heat loss within the asteroid. The peak temperatures might be brought down to ~298K by employing a higher effective bulk asteroid permeability, mimicking a "rubble pile" model.

*Formation Time:* For the canonical set of parameters, but with an accretion time greater than 3.5 Ma, short-lived radionuclides do not generate enough heat to melt any water at all anywhere in the asteroid. An ice fraction of 20% would push this cutoff closer to ~3.75 Ma, but ice fractions smaller than 20% will not provide enough water to produce the observed serpentine contents. Therefore, given our initial heat inputs, final CM parent body accretion occurred no later than ~4 Ma after nucleosynthesis. This result is consistent with estimates from isotopic considerations (Swindle 1993) and other chondrite models (McSween and Bennett 1995, Wilson *et al.* 1999). The later time corresponds to a $^{26}$Al/$^{27}$Al ratio of no less than ~3 * $10^{-6}$ in the accreted material, which is consistent with measurements in chondrules (Russell *et al.*

1996). The earliest accretion time is more loosely bracketed. Accretion at 2 Ma (Runs 7, 8, and 9) results in the highest central temperatures (Fig. 1), but only Run 8, with an initial ice fraction of 40%, avoids temperatures high enough for serpentine dehydration (~530K) (O'Hanley 1996). Runs 7 and 9 get to temperatures over 500K, but this is only in the central regions. Even in these models, there are middle regions of the asteroid where temperature gradients allow water to exist at lower temperatures (Fig. 3). We discuss these regions in more detail in our Discussion section.

*Formation Distance:* The formation distance determines the surface temperature of the asteroid, which in turn determines the depth of the outer layer that never sees liquid water. However, since the surface temperature is never less than 50K, and all runs eventually reach this surface temperature, it has a negligible effect. Volume fractions of the components in Run 1 after 10 Ma are shown in Fig. 5. In the canonical run, the outer 14 km remains ice; in Run 6 (3.5AU), this region is 15.5 km deep. This icy rind persists in all runs and never sees conditions that allow melting of water and subsequent aqueous alteration. The closest to the surface that reactions occur is 6 km in Run 8 (neglecting Runs 7 and 9, whose internal temperatures exceeded 500K where our model is invalid). For a given formation time, a lower ice fraction will push this boundary closer to the surface.

One way to have melting occur in this region is by impact-generated heat that would melt some ice and lead to local, near-surface alteration. Over time, it is possible that the top layers could be altered in this fashion. Impacts might also strip away the icy rind and leave interior, altered material exposed. Another important effect of impacts is that they create a low-porosity, dry, thermally insulating layer over the asteroid. This insulating layer allows heat from decay of long-lived radionuclides to accumulate and may lead to a second, much later, episode of aqueous alteration. The explicit inclusion of impacts in our model is a subject of future work.

*Other parameters:* Within our parameter space, convection and fracturing never occur. This is because these processes depend on large thermal gradients, which are only rarely present in our model.

The choice of alteration reaction determines the amount of heat released during alteration, and this is found to have a large effect on the duration of liquid water. The choice of reaction also has volumetric consequences: the reaction forsterite + 3 water ⇌ antigorite + brucite has a mineral volume change of 45

cm$^3$, while the reaction forsterite + clinoenstatite + 2 water ⇆ antigorite has only a 33 cm$^3$ mineral expansion (Robie 1995). The total volume change, however, is about -10% for both reactions, meaning that the volume initially occupied by water will not quite be filled by the reaction products. If water is the limiting reagent in the equation, then void space is created by these reactions. Examination of altered matrix for expansion textures might help constrain the mode of alteration.

Our model is relatively insensitive to the rate of reaction and the uniformity of material. While including a finite rate of reaction does allow us to better characterize the liquid water phase, the actual rate is so rapid that even varying it by orders of magnitude (such as might be expected due to pH changes (Sverdup 1990)) would produce a small change at the timescales of interest. Similarly, if the starting material were not of small grain size or not uniformly mixed, the effect would be to slow the reaction rate, but this would likely have a negligible effect on the characteristics of the liquid water phase. Additionally, because the reaction time is extremely short compared to any other timescale (e.g. for diffusion of water), and due to the thermal runaway effect (see below), almost every zone of the model shows either full reaction to 60% serpentine (plus some residual olivine and pyroxene, as dictated by the starting assemblage) or no reaction at all. Few zones show intermediate amounts of serpentine.

The model results are sensitive to the solar nebula model used. For instance, we ran a model starting at 240K (a factor of 3 higher than the Cassen (1991) model predicts) but evolved the nebula temperature according to Eq. A2. In this case, essentially the entire asteroid becomes hydrated but the majority of the asteroid subsequently experiences a temperature rise well above the serpentine dehydration point. Clearly, the large range of nebula temperatures that are predicted by models (Wood and Morfill 1988) could be used to produce many different thermal model scenarios. A warmer nebula temperature strongly affects the peak temperature of the asteroid but can still allow the formation of narrow bands within the asteroid which experience conditions consistent with CM observations. However, almost any reasonable choice of nebula temperature, warm or cool, will allow these bands to form, as discussed below.

# Discussion

*Temperature:* The maximum temperature at any point inside the parent body is most sensitive to the time of formation. For a given time of formation, the initial ice fraction dominates the maximum temperature (Fig. 1). This is for three reasons: a) the volume fraction of rock (and thus the power input from radionuclides) increases as the ice fraction decreases; b) the latent heat of fusion of ice serves to take up some of the heat produced by the rock; and c) water has a higher heat capacity than rock and effectively buffers the temperature rise of zones where it is present.

The hydration reaction releases more heat per mole of water used than is required to melt a mole of ice (heat of reaction = 34 kJ per mole of water consumed; latent heat of fusion of water = 6 kJ/mol). This results in a "thermal runaway" that goes on to heat the parent body to a temperature above the melting point of water, once any liquid water is produced at all. Thus, any run where reactions have taken place will produce an internal temperature well over the melting point of water (Fig. 2). All of our runs have central peak temperatures above ~400K, except Run 3 which has not yet reacted after 1 Ma and Run 10 which never melts any water. The peak temperatures might remain closer to the water melting point if the initial ice fraction is more than 40% by volume, a high bulk permeability carries more heat away via diffusion, and/or a later formation time lowers the initial heat input.

There are locations in the asteroid containing liquid water that never gets hotter than the temperature range predicted by CM meteorite observations (273-298K). These occur as relatively thin bands far from the center. The location of this CM-matching zone is dependent mainly on the central temperature; a higher internal temperature pushes water out farther and more rapidly. The width of the zone is partially controlled by permeability, where a higher permeability produces a wider band. Additionally, the bands become narrower the closer to the surface they appear. The widest band in any of our models is produced by Run 3, which had 40% ice. This run produced a band 1 km wide located 28 km from the center of the asteroid. Run 7 produced a band only 4 km from the surface, but it was a mere 0.2 km wide. Thus, a smaller initial ice fraction, a closer accretion distance, an earlier accretion time, or a warmer solar nebula all act to create these bands closer to the surface.

*__Duration__*: Liquid water can exist in the inner asteroid for millions of years (Fig. 3). Since the central region melts first and has the highest peak temperature, the duration of liquid water in the center represents the upper limit of duration for each run. The longest that the liquid phase persists is in Run 4, where a low permeability prevents water from diffusing out for more than 4 Ma. On the other hand, water can be consumed nearly as soon as it is produced, as in Run 2 (Fig. 4), resulting in short ($10^4$ years) durations that roughly agree with the CM meteorite observations. However, these short durations invariably result in a high peak temperature due to the lack of liquid water to further take up heat.

The aqueous phase can persist long after alteration of the original minerals to serpentine is complete. Oxygen isotope exchange is orders of magnitude faster during the hydration reaction than when the altered minerals are in contact with water (Cole and Ohmoto 1986). Since the alteration front is produced as the water melts, and the alteration reaction is extremely rapid, the water actually doing the alteration is quite cool. Thus, the oxygen isotopes of the altered minerals might initially reflect cool temperatures and short times. However, any remaining water would subsequently experience elevated temperatures and persist for millions of years, which might produce effects, such as glass dissolution, which are not observed in CM meteorites.

## Application to Amino Acid Racemization

Amino acids are just one of many organic substances found in CM meteorites (Cronin and Chang 1993). Amino acids in meteorites are particularly interesting, because they are biologically important molecules and because the majority of life on Earth uses exclusively left-handed (laevorotatory, or L) amino acids to carry out life functions rather than their chemically equivalent but mirror-image right-handed (dextrorotatory, or D) counterparts. An intriguing idea is that an enhanced L/D ratio in the prebiotic Earth might have pushed the first biota to prefer left-handed molecules. L-enhancement could occur either by some process on Earth, or by delivery of chirally enhanced extraterrestrial organic material, possibly by way of meteorites or comets (e.g. Chyba and Sagan 1992).

The evolution of amino acids from the time they formed to the time we find them in meteorites, then, is important to our understanding of what might have been deliverable to the early Earth. If an excess of L amino acids was available on meteorite parent bodies, the parent body thermal history is important in determining whether that excess could be preserved in meteorites. Cohen and Chyba (1999) investigated the effects on seven α-hydrogen amino acids of meteorite parent body thermal histories based on CM meteorite observational constraints. Here we describe the evolution of the same suite of amino acids based on our asteroid thermal model results.

There are two main ways amino acids could make their way into meteorite parent bodies. The first is incorporation into the solar nebula of organic molecules observed in the interstellar medium (Pendleton *et al.* 1994). Bonner (1991) reviews the idea that that amino acids produced in the interstellar medium could be non-racemic (D/L≠1) due to physical processes such as parity violation, β-particle bombardment, and exposure to circularly polarized light. The second is through synthesis in liquid water on the parent body itself (Peltzer *et al.* 1984, Schulte and Shock 1998, Schulte 1998). This process requires HCN, which has a limited lifetime in liquid water at 273K of ~$10^4$ years (Peltzer *et al.* 1984). Thus, the amino acids would have been formed within ~$10^4$ years within the parent body, which is rapid compared to our model timescales. However, a mechanism for chiral enhancement during abiotic synthesis is unknown.

Two processes can work on amino acids after they are formed or incorporated into the meteorite parent body: racemization and thermal decomposition. Racemization is a natural, exponential process by which amino acids change handedness, from D to L and vice versa, until a given mixture approaches a D/L ratio of 1. This process is most effective in solution, and proceeds faster at higher temperatures. Thermal decomposition occurs at temperatures high enough to destabilize the molecule entirely. Using the racemization parameters and model of Cohen and Chyba (1999) and the thermal decomposition parameters of Rodante (1992) we evolved the amino acids alanine, aspartic acid, and isoleucine through the temperature-time histories generated by the thermal model. We began with a D/L ratio of 0.9, which is roughly the excess Cronin and Pizzarello (1997) report in racemization-resistant amino acids. Racemization rates in solution were used when liquid water was present and dry rates were used when either only ice was present or no water at all existed.

Cool (273-298K) water only exists for a short time ($10^3$-$10^4$ years) anywhere within the asteroid. The aqueous phase begins cool, but the water is quickly heated by the exothermic hydration reaction. This hot water can persist for a few millions of years and commonly reaches temperatures up to 400K due to overpressure. At cool temperatures, significant racemization usually takes a longer time than that predicted by the model results, but in some cases, partial racemization occurs for some amino acids in the suite and not for others. If all the amino acids had the same original D/L ratio, then finding different D/L ratios among this suite could help constrain the thermal history undergone by these amino acids. On the other hand, the entire suite of amino acids is completely racemized within a few thousand years when contained within a 350-400K solution. In regions of the parent body that contained a hot solution, amino acids would quickly lose any trace of an initial chiral signature. In model zones that become dry and continue to heat up, thermal destruction becomes important. For the amino acids considered here, thermal destruction in the solid state is more than 90% complete at ~473K (Rodante 1992). In regions where deserpentinization occurs, temperatures well exceed the thermal destruction temperature and no amino acids would be expected to be found at all.

The bands of the model asteroid which could produce CM-like meteorites are defined as zones which never get hotter than 298K but still experience hydration. In these bands, the temperature could remain low because the water was either completely consumed by the hydration reaction or was frozen out by the cold temperatures near the surface. In both cases, the duration of the cool water phase was extremely short, again, on the order of $10^3$ years. In these zones, little to no significant racemization occurs, implying that the D/L ratios observed in CM meteorites are near or identical to their initial values. Throughout the rest of the model asteroid, however, either complete racemization or thermal destruction is predicted to occur.

# Conclusions

The evolution of amino acids on meteorite parent bodies is extremely sensitive to the temperature and duration of the liquid water phase. Therefore, our model attempts to characterize this phase in detail for CM type parent bodies.

The conditions for reaching the melting point of water within our model depend on the accretion time since nucleosynthesis, since the predominant energy source for this temperature rise is from radioactive decay (heat of accretion only adds 50K, and we ignore magnetic induction and impact-generated heat). For an asteroid at 3 AU, a formation time after ~4 Ma after nucleosynthesis means that so much radionuclide decay has gone on already that the asteroid never reaches the melting point of water.

In scenarios where water does melt, hydration reactions are critical in producing a thermal runaway. If the total reaction heat from all reactions going on during the aqueous phase is more than the heat of fusion of water (6 kJ) per mole of water consumed in the hydration reaction, then a runaway melting of water will ensue. All the reactions we considered produced enough heat for this runaway scenario to occur. Thus, if reactions do not occur, $T_{max}$<273K, but if they do, $T_{max}$>~460K. A set of hydration reactions that is only slightly exothermic could be imagined if phyllosilicate formation is a two-step process, occurring partially in the solar nebula and finishing in situ. However, the water fugacity in the nebula was probably too low to support aqueous reactions.

The temperature and duration of the liquid water phase is radially variable, ranging from hot temperatures in the center of the asteroid, through narrow bands of cool aqueous conditions, to always-frozen rinds. The mineral assemblage produced, therefore, is radially variable as well (Fig. 5). It has been suggested (Zolensky *et al*. 1989) that both CI- and CM-type material could be formed by progressively altering the same initial assemblage, perhaps by this kind of parent-body history.

While it is useful to attempt to match thermal model results with CM meteorite observations, thus far our model can only produce narrow (~1km) bands of the asteroid that experience a cool aqueous phase for a short time. Mechanisms for lowering the temperatures and durations of the aqueous phase in these models seem to be a) increased bulk permeability, b) larger fraction of ice, and c) lower heat of reaction. On the other hand, incorporating impact-generated heat or a warmer solar nebula could help produce these conditions in the exterior of the parent body without regard to the interior conditions.

The model constraints on the characteristics of the liquid water phase are well defined no matter what initial parameters are chosen. A cool solution exists for a short time ($10^3$ years), then is either consumed by reactions, frozen out, or heated up rapidly to high temperatures. In regions where cool water exists for short times, amino acids would be expected to survive and largely retain their initial D/L ratio.

Conversely, the majority of the model asteroid experiences conditions that completely racemize or even destroy amino acids.

# Appendix

Here we present the numerical and mathematical details of the model used to calculate the internal evolution of carbonaceous chondrite parent bodies. All units are SI (mks) unless otherwise noted.

### A. Heat Flow

We use a one-dimensional implicit finite-difference method with 500 zones of equal size to solve the spherically symmetric heat-conduction equation in the presence of internal heating:

$$\frac{dT}{dt} = \frac{1}{r^2} \frac{d}{dr}\left[r^2 \kappa \frac{dT}{dr}\right] + \frac{Q}{c_p} \quad (A1)$$

where T is the temperature, r is the radius, $\kappa$ is the thermal diffusivity, $c_p$ is the specific heat, and Q is the power input per unit mass due to radionuclide decay (given by Eq. A5). We use an internal boundary condition that sets dT/dr=0 at the origin and a surface boundary condition based on Cassen (1994) given by:

$$T(r = R) = \frac{860}{D^{1.13} t^{1.1}} \quad (A2)$$

where R is the radius of the asteroid, D is the distance in AU of the asteroid from the sun and t is the time in millions of years since the collapse of the nebula. Eq. A2 assumes that the surface of the asteroid is always in equilibrium with its surroundings. The surface temperature is set by the value of Eq. A2, or 50K, whichever is larger. For our canonical run (D=3, t=3), this is about 75K. However, for the entire asteroid, the initial temperature, $T_i$, is given by

$$T_i = \sqrt{T(r = R)^2 + T_{acc}^2} \quad (A3)$$

where the accretion temperature, $T_{acc}$, is derived from converting all of the asteroid's initial potential energy to thermal energy, and is given by:

$$T_{acc} = \frac{16}{3} \frac{\pi^2 \bar{\rho}^2 R^5 G}{M \bar{c}_p} \quad (A4)$$

where M is the total mass of the asteroid and $\bar{\rho}$ and $\bar{c}_p$ are the density and specific heat, respectively, averaged over the whole asteroid. For our canonical run with a 100-km diameter asteroid, $T_i = 90K$. Since $\bar{\rho}$ and $\bar{c}_p$ depend on temperature, some iterating is required to get a self-consistent initial temperature. We use volume weighting to calculate k, the thermal conductivity, and ρ, the density, for a given zone while we use mass weighting for $c_p$; then, κ is given by $k\rho^{-1}c_p^{-1}$.

For a 50-km radius asteroid accreting material from infinity, the potential energy input is about $10^{23}$ J, enough to raise the material temperature about 50K. On the other hand, short-lived radionuclides contribute about two orders of magnitude more energy over the first 1 Ma. We use average CM radionuclide abundances from Mason (1971) for a suite of 16 long- and short-lived radionuclides (Table AI), and the same concentration is used for the olivine, pyroxene, and non-reactive rock components. We began with a $^{26}Al/^{27}Al$ ratio of $5.06*10^{-5}$, the well-established upper limit of $^{26}Al$ abundance in CAI's, and the inferred initial value of CM-type refractory objects (MacPherson *et al.* 1995). We obtained the abundance of other nuclides in CM meteorites (Mason 1971) and calculated the mass fraction $A_\ell$ at 4.55 Ga using a typical chondritic Si abundance of 10.3% by weight (Anders and Grevesse 1989). Using decay constants and energies found in Lide (1994) and in Faure (1986), we calculated the total power contribution $Q_\ell$ from each radionuclide.

**TABLE AI**
**Radionuclides included in the model**

| Element $\ell$ | Mass fraction A | Decay constant (yr$^{-1}$) λ | Decay energy (J) | Power (W/kg) Q |
|---|---|---|---|---|
| $^{26}Al$ | 5.06×10$^{-7}$ | 1.02×10$^6$ | 4.005×10$^{-13}$ | 1.45×10$^{-7}$ |
| $^{40}K$ | 7.93×10$^{-7}$ | 1.82×10$^9$ | 1.32×10$^{-13}$ | 2.75×10$^{-11}$ |
| $^{41}Ca$ | 1.37×10$^{-10}$ | 1.49×10$^5$ | 4.21×10$^{-14}$ | 1.80×10$^{-11}$ |
| $^{53}Mn$ | 6.70×10$^{-10}$ | 5.34×10$^6$ | 5.96×10$^{-14}$ | 2.70×10$^{-12}$ |
| $^{87}Rb$ | 5.17×10$^{-7}$ | 7.04×10$^{10}$ | 2.73×10$^{-14}$ | 4.41×10$^{-14}$ |
| $^{107}Pd$ | 1.49×10$^{-10}$ | 9.38×10$^6$ | 3×10$^{-15}$ | 8.50×10$^{-15}$ |
| $^{129}I$ | 3.94×10$^{-12}$ | 2.45×10$^7$ | 1.91×10$^{-14}$ | 4.55×10$^{-16}$ |
| $^{146}Sm$ | 9.53×10$^{-11}$ | 1.49×10$^8$ | 2.23×10$^{-13}$ | 1.87×10$^{-14}$ |
| $^{147}Sm$ | 3.19×10$^{-8}$ | 1.53×10$^{11}$ | 2.5×10$^{-13}$ | 6.78×10$^{-15}$ |
| $^{176}Lu$ | 8.83×10$^{-9}$ | 5.48×10$^{10}$ | 1.188×10$^{-13}$ | 2.08×10$^{-15}$ |
| $^{187}Re$ | 3.45×10$^{-8}$ | 6.06×10$^{10}$ | 2.5×10$^{-16}$ | 1.45×10$^{-17}$ |
| $^{232}Th$ | 5.53×10$^{-8}$ | 1.50×10$^{10}$ | 3.98×10$^{-12}$ | 1.21×10$^{-12}$ |
| $^{235}U$ | 7.60×10$^{-9}$ | 1.02×10$^9$ | 4.52×10$^{-12}$ | 2.75×10$^{-12}$ |
| $^{238}U$ | 2.44×10$^{-8}$ | 6.43×10$^9$ | 4.74×10$^{-12}$ | 1.45×10$^{-12}$ |
| $^{244}Pu$ | 1.75×10$^{-10}$ | 1.18×10$^8$ | 4.665×10$^{-13}$ | 5.41×10$^{-14}$ |
| $^{247}Cm$ | 3.20×10$^{-11}$ | 2.25×10$^7$ | 5.352×10$^{-13}$ | 5.89×10$^{-14}$ |

Total power
1.455×10⁻⁷

The total heat production due to the decay of radionuclides is given by

$$Q = m_{rock} \sum_\ell A_\ell Q_\ell e^{-\lambda_\ell t} \quad (A5)$$

where $A_\ell$, $Q_\ell$, and $\lambda_\ell$ are given in Table AI. The presence of $H_2O$ requires that Q be scaled by $m_{rock}$, the fraction of the initial total mass that is not $H_2O$.

The power supplied by radionuclides at time=0 (defined as 4.55Ga ago) is $1.455*10^{-7}$ W kg$^{-1}$. This compares to $6.7*10^{-9}$ W kg$^{-1}$ determined by Grimm and McSween (1989) to be the minimum amount of power necessary to melt water on a 100-km CM type body. The $^{26}Al/^{27}Al$ ratio to generate this minimum power was $1.6*10^{-6}$, a factor of three different from our value.

**B. $H_2O$ Phase Transitions**

Since the lithostatic pressure for a 100-km diameter asteroid is less than a kilobar, we assume that the melting point of ice occurs at $T_{melt}$=273.15K, independent of pressure. However, due to the likely presence of salts, we use $T_{freeze}$=268K, the $H_2O$-$MgSO_4$ eutectic point (Kargel 1998). We chose $MgSO_4$ because it is quite common. The recent discovery of large amounts of NaCl in some chondrites (Zolensky 1999) may argue for the inclusion of this salt as well, but the difference in effect of any salt or salt combination is probably very small.

Liquid-water phase transitions are tracked using a modified version of the algorithm of Reynolds *et al.* (1966) which uses a partial melt state. For example, if, for a particular time step, Eq. A1 results in a temperature rise such that T is slightly above $T_{melt}$, then, in the presence of ice $H_2O$, the temperature rise above $T_{melt}$ is suppressed and a fraction, $\Delta m_j$ of the mass in ice $H_2O$ is converted to liquid:

$$\Delta m_j = \frac{(T - T_{melt}) c_p}{H_f} \quad (A6)$$

where $H_f$ is the latent heat of fusion of ice. Similar relations apply for the freezing of liquid $H_2O$.

For $H_2O$ vapor, the Clausius-Clapeyron relation is assumed to hold so that

- 16 -

$$P_{vap} = P_0 e^{-\left(T_0/T\right)} \quad (A7)$$

where, following Grimm and McSween (1989), $P_0 = 3.58 \times 10^{12}$ Pa and $T_0 = 6140$K in the presence of ice and $P_0 = 4.70 \times 10^{10}$ Pa and $T_0 = 4960$K in the presence of liquid $H_2O$. In the presence of mixed phases, a mass-weighted average of $P_{vap}$ is used. Since T < 500K for these models, an ideal equation of state is assumed. Thus, $P_{vap} V_{void} = n R_G T$ where $V_{void}$ is the total void space volume, n is the number of moles of $H_2O$ vapor, and $R_G$ is the universal gas constant. Thus, $M_{vap}$, the mass of $H_2O$ gas present in a given time step and zone, is

$$M_{vap} = \frac{0.018 P_{vap} V_{void}}{R_G T} \quad (A8)$$

It is further assumed that $H_2O$ vapor fills the pore spaces of a given zone completely, immediately reaching equilibrium, with a corresponding change in temperature, $\Delta T$, given by:

$$\Delta T_j = \frac{\Delta M_{vap} H_v}{M_j c_p} \quad (A9)$$

where $M_j$ is the total mass in the jth zone and $H_v$ is the latent heat of vaporization (or sublimation as appropriate). The mass of gas either vaporizing or condensing, $\Delta M_{vap}$ is simply $M_{vap}$ minus $M_{vap}$ from the previous time step. For vaporization, it is assumed that all liquid $H_2O$ evaporates before any solid $H_2O$ sublimes. The relation for $M_{vap}$ is limited by the amount of liquid or solid $H_2O$ present. For the heat of vaporization, we use a 3rd order polynomial fit to data given in Lide (1994):

$$H_v = 3713997.2 - 7822.6569T + 17.613373T^2 - 0.019018061T^3 \quad (A10)$$

where, due to limited data in Lide (1994), T is capped at 400K.

**C. Vapor Diffusion**

In the absence of gravity, the diffusion of $H_2O$ vapor through the asteroid is assumed to obey Darcy's law (see, e.g., Turcotte and Schubert 1982):

$$u = -\frac{K_{vap}}{\eta} \frac{dP_{vap}}{dr} \quad (A11)$$

where u is the vertical velocity of the gas (evaluated between zones), $K_{vap}$ is the effective permeability of the rock to vapor diffusion scaled from the 20%-porosity value $K_0$, and $\eta$ is the viscosity of the gas. Thus, the mass of vapor transported is

$$\Delta M_{vap} = 4\pi r^2 \rho_{vap} u \quad (A12)$$

where the density of the water vapor, $\rho_{vap} = M_{vap} V_{void}^{-1}$. This results in either upward or downward motion of water vapor, depending on the sign of the pressure gradient. The pressure of the water vapor, $P_{vap}$ now out of thermal equilibrium, is recalculated.

At low temperatures, vapor pressure becomes low enough that the path length of vapor-vapor collisions becomes longer than that of vapor-rock collisions. The vapor molecules can then "stick" to the rock surface and create a net flow, called Knudsen flow. This results in an effective increase in the permeability. Following Grimm and McSween (1989), and assuming a pore size of 50 μm, we calculate the temperature and pressure at which Knudsen flow begins. The temperature is approximately 290K and the pressure is given by

$$P_{Knud} = \frac{R_G T}{4\sqrt{2}\pi N b^2 d} \quad (A13)$$

where N is Avogadro's number, b is the radius of the $H_2O$ molecule, and d is the pore size. Thus, if T < 290K and $P_{vap} < P_{Knud}$, then

$$K_{vap} = K_0 \frac{v_{void}}{0.2} \frac{P_{Knud}}{P_{vap}} \quad (A14)$$

where $v_{void}$ is the volume fraction of void space. Otherwise,

$$K_{vap} = K_0 \frac{v_{void}}{0.2} \quad (A15)$$

This assumes that $H_2O$ vapor and liquid are immiscible fluids. Similar to the mass transport, the diffusing vapor carries heat with it, resulting in a change in temperature of

$$\Delta T = \frac{3}{2} N \frac{0.018 \, \Delta M_{vap} k_b T}{c_p M_j} \quad (A16)$$

where $k_b$ is the Boltzmann constant and $M_j$ is the total mass in the zone. The diffusion of water vapor is stopped if the gas is blocked in by a zone with little pore space (for numerical reasons, the minimum void space fraction, $v_{void}^{min}$, is set to $10^{-6}$), if a given zone wants to diffuse in both directions (in which case it is only allowed to diffuse upward), or if the pressure gradient is extremely small ($10^{-6}$ $P_{vap}$).

D. Venting

It is assumed that all of the water vapor in the topmost zone is lost to space every time step. This venting can be appreciable over 4.55 Ga. However, internal gas venting also occurs when, due to evaporation and diffusion, $P_{vap} > P_{lith} + \tau$ where $\tau$ is the tensile strength of the rock (assumed to be $10^7$ Pa) and the lithostatic pressure, $P_{lith}$, which is approximated by

When gas venting occurs, enough vapor (such that $P_{vap} = P_{lith} + \tau$) is transported upward one zone. This transport is analogous to instantaneous diffusion. Similarly, if $v_{void} < P_{hyd} = B\left(\dfrac{v_{new}}{v_{pore}} - 1\right)$ (A19), all vapor in that zone is transported upwards. For some models, this results in explosive venting for a thickness of a few zones; the venting never reaches the surface.

E. Fracturing

If diffusion and condensation or the freezing and subsequent expansion of $H_2O$ results in a volume fraction of solid and/or liquid $H_2O$, $v_{new}$, greater than the available pore space fraction, $v_{pore}$, we might have hydraulic fracturing. This occurs when the hydraulic pressure, $P_{hyd}$, exceeds the confinement pressure:

$$P_{hyd} > P_{lith} + \tau \quad (A18)$$

where

$$P_{hyd} = B\left(\dfrac{v_{new}}{v_{pore} - 1}\right) \quad (A19)$$

and B is the bulk modulus of liquid or solid $H_2O$, as appropriate. If fracturing occurs, enough mass is transported upwards so that Eq. A18 becomes an equality. For the models presented here, fracturing never occurs and the only movement of mass between zones is that due to diffusion or venting of $H_2O$ vapor.

However, some models do run out of void space. When this happens, we assume the rock is incompressible but compress the liquid and solid $H_2O$ such that they exactly fill the original pore space.

This way of handling of diffusion, venting, and fracturing is not fully self-consistent in terms of the connectivity of the rock. That is, we assume full connectivity of the pore spaces within each zone for vapor equilibrium and the movement of liquid $H_2O$, moderate connectivity between zones (determined by the permeability) for vapor diffusion, and zero connectivity between zones for liquid and solid $H_2O$.

F. Hydrothermal Convection

Following Grimm and McSween (1989), and references therein, we assume hydrothermal convection occurs if the Rayleigh number, Ra, exceeds $4\pi^2$. Ra is approximately given by

$$Ra = K_0 \rho_{liq}^2 \frac{v_{liq}}{v_{liq} + v_{void}} \left(\tfrac{4}{3}\pi G \bar{\rho} r\right) \frac{\alpha_{liq} c_p^{liq} \Delta T \Delta r}{\eta_{liq} k} \quad (A20)$$

where $\eta_{liq}$ is the viscosity of water, k is the volume-fraction averaged thermal conductivity of the zone, and $\Delta T$ is the temperature drop across one zone of size $\Delta r$, and $\alpha_{liq}$ is the temperature-dependent coefficient of thermal expansion of water, given by

$$\alpha = -\frac{1}{\rho_{liq}} \frac{d\rho_{liq}}{dT} \quad (A21)$$

where $\rho_{liq}$ is given by Eq. A39. Convection results in an effective increase in the thermal diffusivity, $\kappa$, so that for $4\pi^2 < Ra < 8\pi^2$,

$$\kappa = \left(\frac{Ra}{4\pi^2}\right)^{1.3} \kappa_0 \quad (A22a)$$

and for $Ra > 8\pi^2$,

$$\kappa = 1.6 \left(\frac{Ra}{4\pi^2}\right)^{0.6} \kappa_0 \quad (A22b)$$

where $\kappa_0 = k\, c_p^{-1}\, \rho^{-1}$, the thermal diffusivity in the absence of convection. However, for the models presented here, we find that convection never occurs.

G. Hydration Reactions

There are a number of alteration reactions that can be put into the model; however, rate of reaction information exists only for the simple reaction

$$2\,Mg_2SiO_4\,(\text{forsterite}) + 3\,H_2O \rightleftarrows Mg_3Si_2O_5(OH)_4\,(\text{antigorite}) + Mg(OH)_2\,(\text{brucite}) \quad (A23)$$

(Wegner and Ernst 1983), with a total heat of reaction of 367 kJ per mole of serpentine produced. This reaction is clearly not the dominant one for CM meteorites, because brucite does not exist in equal amounts to the serpentine-group minerals. Therefore, we chose as the dominant reaction

$$Mg_2SiO_4\,(\text{forsterite}) + MgSiO_3\,(\text{enstatite}) + 2\,H_2O \rightleftarrows Mg_3Si_2O_5(OH)_4\,(\text{antigorite}) \quad (A24)$$

This reaction releases 69 kJ per mole of serpentine produced (Robie 1995) and results in a net increase in void space of approximately 3 cm³ per mole of serpentine produced. The reaction proceeds only in the presence of liquid $H_2O$.

Since kinetic data does not exist for Eq. A24, we apply the rate of reaction of Eq. A23 given by Wegner and Ernst (1983). We assume that the rate of reaction is linear with pressure and that it follows an Arrhenius form. Thus, fitting Wegner and Ernst's (1983) 1 kbar high temperature data points for a grain size of 37-62 μm, we get a rate constant, $K_r$, in moles of reactions per year:

$$K_r = 4383\,P\,e^{-\frac{3463}{T}} \quad (A25)$$

where $P = P_{lith} + P_{vap}$ in kbars. In addition, we assume that the length scale of the reaction (set by the ratio of volume to surface area for the reacting rock) is temporally and spatially constant. Finally, we get the number of moles of serpentine produced, $\Delta m_{serp}$, in a given time step, $\Delta t$ (in years), of

$$\Delta m_{serp} = m_{avail}\left(1 - e^{-K_r \Delta t}\right) \quad (A26)$$

where $m_{avail}$ refers to the limiting reagent, either $m_{olivine}$ or $2m_{liquid}$. Once liquid water becomes available, the limiting reagent is olivine, as we choose a volume of pyroxene to be present such that both minerals are consumed essentially equally. In models where all the liquid is completely consumed, the liquid becomes the limiting reagent at the end of the aqueous alteration phase.

Since the reaction is exothermic, the temperature of the zone is increased in a method analogous to that done when the $H_2O$ vapor condenses. That is,

$$\Delta T = \frac{H_r \Delta m_{serp}}{c_p M_j} \quad (A27)$$

where $H_r$ is the heat of reaction of 69 kJ/mole (Robie 1995). The dehydration temperature is approximately 530K (O'Hanley 1996). Since the maximum temperature obtained in the majority of the models presented here is less than 500K, and some of our thermodynamic values break down above 500K, it is assumed that no dehydration reactions occur. Finally, as mentioned above, we assume, within a given zone, full pore space connectivity so that any liquid $H_2O$ present will react, if possible.

**H. Material Properties**

The asteroid is characterized by seven separate components: non-reactive rock, forsterite, enstatite, serpentine, liquid $H_2O$, solid $H_2O$, and the pore space (filled with $H_2O$ vapor). Void space is set to an initial volumetric value of ~16% to try to match the final CM bulk porosity of 20% (Britt and Consolmagno 1997). The thermal conductivity, K, the specific heat, $c_p$, the density, $\rho$, and, for $H_2O$, the viscosity, need to be characterized.

**a. Thermal Conductivity**

Except for $H_2O$, the thermal conductivity is assumed to be independent of temperature. Thermal conductivity is dominated by lattice conductivity; radiative conductivity is thought to be important only above 500K (Schatz and Simmons 1972). For forsterite and serpentine, we use mineral values of 5.155 and 2.95 $Wm^{-1}K^{-1}$ respectively (Touloukian *et al.* 1970a). We also use the forsterite value for enstatite. We use data from Horai and Susaki (1989) to determine an approximate 'zero-porosity' thermal conductivity for carbonaceous chondrites, scale it for porosity by $\exp(-1.9/(1-v_{void}))$ and use the resulting value, 2.8 $Wm^{-1}K^{-1}$, for the thermal conductivity for the non-reactive rock. For solid $H_2O$, we use (Yen 1981):

$$k_{ice} = 9.828 \exp(-0.0057T) \quad (A28)$$

For $H_2O$ vapor, we use (Touloukian *et al.* 1970a):

$$k_{vap} = -0.0143 + 0.000102T \quad (A29)$$

This expression is clearly invalid for low T so, lacking any other data, we set T to be $T_j$ or 150K, whichever is greater. For liquid $H_2O$, we use matched expressions based on Touloukian *et al.* (1970a):

$$k_{liq} = -0.581 + 6.34*10^{-3}T - 7.93*10^{-6} T^2 \quad (T < 410K) \quad (A30)$$

$$k_{liq} = 0.9721 (-0.142 + 4.12*10^{-3} T - 5.01*10^{-6} T^2) \quad (T > 410K) \quad (A30a)$$

**b. Specific Heat**

The specific heat for all components is temperature dependent. For forsterite, enstatite, and liquid $H_2O$ we fit a 3rd order polynomial to data from a variety of sources (Touloukian *et al.* 1970b, Barin and Knacke 1973, Chase 1985, Knacke et al 1991, and Lide 1994) and get:

$$\log c_p (\text{forsterite}) = -11.32 + 13.58x - 4.25 x^2 + 0.44 x^3 \quad (A31)$$

$$\log c_p (\text{enstatite}) = -8.62 + 10.39x - 3.00 x^2 + 0.28 x^3 \quad (A32)$$

$$\log c_p (\text{water}) = 8.25 - 4.18x + 1.12 x^2 - 0.076 x^3 \quad (A33)$$

where $x \equiv \log T$. These expressions, good to ~20%, are valid from approximately 50 to 500K. The value for forsterite is used for the non-reactive minerals as well. For antigorite, we use, for T > 273K (Barin 1977),

$$c_p (\text{antigorite}) = 1145 + 0.048T - 2.65*10^7 T^{-2} \quad (A34a)$$

while for T < 273, we use an expression derived from data on talc (Touloukian *et al.* 1970b) and match to the above expression to get

$$\log c_p (\text{antigorite}) = -0.59 - 1.51x + 2.82x^2 - 0.66x^3 \quad (A34b)$$

where again, $x \equiv \log T$.

For solid $H_2O$, we use values from Yen (1981):

$$c_p (\text{ice}) = -49.97 + 9.5T \quad (50 < T < 95K) \quad (A35a)$$

$$= 126.89 + 7.5T \quad (95 < T < 150K) \quad (A35b)$$

$$= 152.46 + 7.12T \quad (T > 150K) \quad (A35c)$$

For $H_2O$ vapor, we use a linear fit to data in Lide (1994) and get:

$$c_p (\text{vapor}) = 1730.54 + 0.45T \quad (A36)$$

Ideally, $c_v$, not $c_p$, should be used in Eq. A1. However, we use $c_p$ because it has been better determined and, where both have been measured, differs from $c_v$ by only ten percent at low temperatures (Ghosh and McSween 1998).

### c. Density

The densities for forsterite and enstatite are 3210 and 3190 kg m$^{-3}$ respectively (Lide 1994). We use a density for serpentine of 2470 kg m$^{-3}$, derived from the volume ratios given in Grimm and McSween (1989). For the non-reactive minerals, we use a density of 3630 kg m$^{-3}$, from the carbonaceous chondrite value in Horai and Suzuki (1989) extrapolated to zero porosity. For $H_2O$ vapor, we use the ideal gas law as discussed above:

$$\rho_{vap} = P_{vap} \, 0.018 \, T^{-1} \, R^{-1} \quad (A37)$$

For $H_2O$ liquid, we use (Keyes 1928):

$$\rho_{liq} = -221 + 13.1T - 0.0507T^2 + 8.49E{-}5\,T^3 - 5.48E{-}8\,T^4 \quad (A38)$$

while for solid $H_2O$ we use (Yen 1981)

$$\rho_{ice} = -1.32x + 935.32 \quad (T < 137K) \quad (A39a)$$

$$\rho_{ice} = -46.9x + 1032.71 \quad (T > 137K) \quad (A39b)$$

If there is any compression due to running out of void space, the density of $H_2O$ liquid or solid is increased proportionately.

### d. Viscosity

The viscosity of liquid $H_2O$ is required for convection calculations while the viscosity of $H_2O$ vapor is required for diffusion. We take both expressions from Grimm and McSween (1989):

$$\eta_{liq} = 10^{\left(\frac{247.8}{T-140}\right) - 4.6} \quad (A40)$$

$$\eta_{vap} = 8.04*10^{-6} + 4.07*10^{-8} \, T \quad (A41)$$

**TABLE AII**
**List of variables and constants**

| | | | |
|---|---|---|---|
| $A_\ell$ | Mass fraction of radionuclide $\ell$, from | $P_0$ | Pressure at infinite temperature (Pa) |

Table I

| | | | |
|---|---|---|---|
| $\alpha$ | Coefficient of thermal expansion of liquid water ($K^{-1}$) | $P_{hyd}$ | Hydraulic pressure (Pa) |
| B | Bulk modulus of $H_2O$=2.29*$10^9$ Pa (liquid) or $10^{10}$ Pa (ice) | $P_{Knud}$ | Pressure at which Knudsen flow begins |
| b | Radius of an $H_2O$ molecule=5*$10^{-11}$ m | $P_{lith}$ | Lithostatic pressure (Pa) |
| $c_p$ | Heat capacity (J $kg^{-1}$ $K^{-1}$) | $P_{vap}$ | Pressure of water vapor (Pa) |
| $\bar{c}_p$ | Heat capacity averaged over the entire asteroid (J $kg^{-1}$ $K^{-1}$) | Q | Power from radionuclide decay (W $kg^{-1}$) |
| D | Distance from sun (AU) | R | Radius of the asteroid=50,000 m |
| d | Pore size=5*$10^{-5}$ m | Ra | Rayleigh number |
| G | Gravitational constant=6.67 * $10^{-11}$ N $m^2$ $kg^{-2}$ | $R_G$ | Gas constant=8.314 J $mol^{-1}$ $K^{-1}$ |
| $H_f$ | Latent heat of fusion of water=3.3*$10^5$ J $kg^{-1}$ | r | Radius (m) |
| $H_v$ | Latent heat of vaporization of water (J $kg^{-1}$) | $\rho$ | Density (kg $m^{-3}$) |
| $\eta_{liq}$ | Viscosity of liquid water (Pa s) | $\bar{\rho}$ | Density, averaged over the asteroid (kg $m^{-3}$) |
| $\eta_{vap}$ | Viscosity of water vapor (Pa s) | T | Temperature (K) |
| j | Zone index | $T_0$ | Empirical e-folding temperature (K) |
| $K_0$ | Permeability of rock to vapor diffusion ($m^2$) | $T_{acc}$ | Accretion temperature (K) |
| $K_r$ | Rate constant of hydration reaction (mol $yr^{-1}$) | $T_i$ | Initial temperature (K) |
| $K_{vap}$ | Permeability of rock to vapor diffusion, scaled by void space and Knudsen flow ($m^2$) | $T_{freeze}$ | Freezing temperature of $H_2O$-$MgSO_4$ eutectic=268.15 K |
| k | Thermal conductivity (W $m^{-1}$ $K^{-1}$) | $T_{melt}$ | Melting temperature of pure water=273.15K |
| $k_b$ | Boltzmann's constant=1.38*$10^{-23}$ J $K^{-1}$ | t | Time (s) |
| $\kappa$ | Thermal diffusivity ($m^2$ $s^{-1}$) | $\tau$ | Tensile strength of rock = $10^7$ Pa |
| $\ell$ | Radionuclide index | u | Darcy velocity of water vapor (m $s^{-1}$) |
| $\lambda_\ell$ | Radionuclide decay constant ($yr^{-1}$) | $V_{void}$ | Volume of void space ($m^3$) |
| $\Delta M_{vap}$ | Change in mass between zones due to vapor diffusion (kg) | $v_{liq}$ | Volume fraction of liquid |
| m | Mass fraction | $v_{pore}$ | Volume fraction of pore space |
| $m$ | Moles | $v_{vap}$ | Volume fraction of water vapor |
| N | Avogadro's number=6.02 * $10^{23}$ atoms $mol^{-1}$ | $v_{void}$ | Volume fraction of void space |

# Acknowledgements

We thank Joe Plassman for helping us with computing resources and Fulvio Melia for not hindering us. M. Zolensky and G. McDonald provided useful reviews. We also gratefully acknowledge encouragement from and discussions with A. Ghosh, P. Cassen, D. Woolum, and H. McSween. B.A.C. is supported by a NASA Space Grant Fellowship and R.F.C. by a NASA GSRP Fellowship. This research has made use of NASA's Astrophysics Data System Abstract Service.

- 29 -

**Figure Captions**

**FIG. 1.** Temperature (K) versus time after nebula collapse (Ma) at the center of the model asteroid for the 10 runs. Run 8 was terminated after all water either reacted or froze out. Run 5, due to its large Darcy velocity, was the most computationally intensive calculation and was thus terminated once the duration of liquid water had been fairly well constrained (see Figure 3). The legend is common to figures 1-4.

**FIG. 2.** Temperature profile throughout the model asteroid 1Ma after accretion. The accretion time is different for each model run. The importance of the surface temperature, which is based on the nebula model, is clearly shown. Note that Run 3 has not yet reacted while Run 10 never will; thus, both profiles are at less than 273 K.

**FIG. 3.** Duration of the liquid water phase (years) versus radius (km) throughout the model asteroid. Except for narrow rinds near the surface, liquid water either never exists or exists for about 1 Ma. The exception is Run 2 which, due to its low initial ice fraction, consumes most of its ice, reaches a higher central temperature (see Figure 1), and thus pushes the remainder of its liquid water to larger radii.

**FIG. 4.** Duration and temperature of liquid water in the center of the model asteroid. In general, a higher central temperature results in a shorter duration of the liquid water phase unless the initial ice fraction is large (e.g. Run 8) or the permeability is low (e.g. Run 4).

**FIG. 5.** Final volume fractions of ice, void space, hydrous minerals, and anhydrous minerals versus radius (km) throughout the model asteroid for Run 1. No hydrous minerals are formed within ~14 km of the surface. Any residual water ice resides in the upper ~30 km of the asteroid.

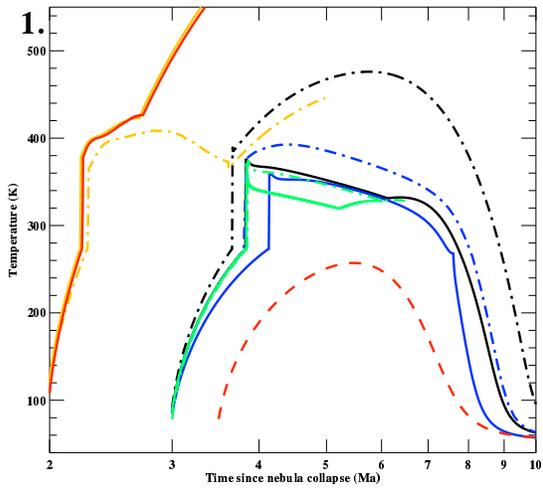
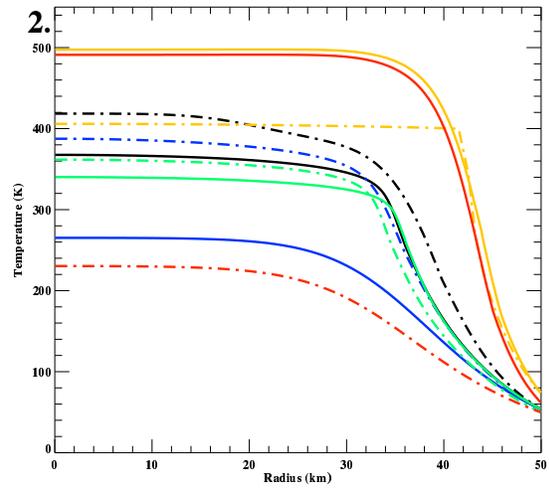
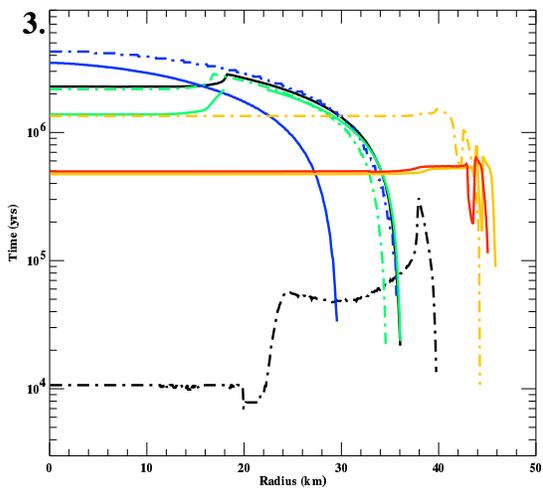
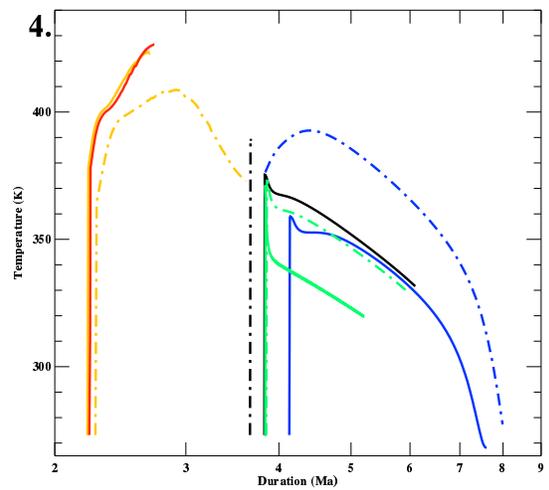

**Legend for Figures 1-4**

| | |
|---|---|
| ——— | Run 1 |
| —·—·— | Run 2 |
| ——— | Run 3 |
| —·—·— | Run 4 |
| ——— | Run 5 |
| —·—·— | Run 6 |
| ——— | Run 7 |
| —·—·— | Run 8 |
| ——— | Run 9 |
| —·—·— | Run 10 |

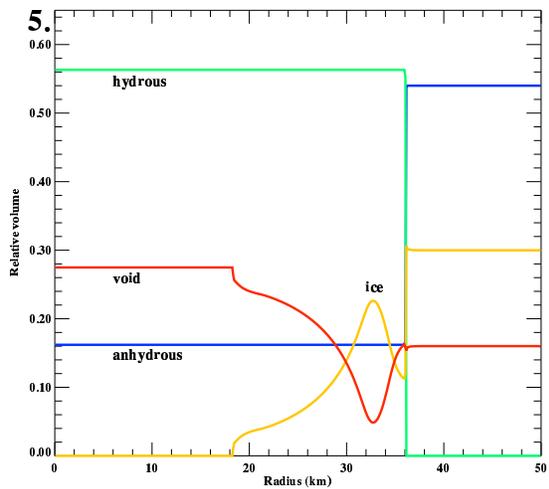